\documentclass[preprint,showpacs,showkeys,aps]{revtex4}
\usepackage{epsf}
\usepackage{dcolumn}
\usepackage{bm}
\everymath{\displaystyle}
\begin{document} 

\title{Potential for measurement of the tensor magnetic polarizability of
the deuteron in storage ring experiments}

\author{\firstname{Alexander J.}~\surname{Silenko}} 
\affiliation{Institute of Nuclear Problems, Belarusian State
University, Minsk 220030, Belarus}

\date{\today}

\begin {abstract}
General formulas describing deuteron spin dynamics in storage
rings with allowance for the tensor electric and magnetic
polarizabilities are derived. It is found that an initially
tensor-polarized deuteron beam can acquire a final horizontal
vector polarization of the order of 1\%. This effect allows one to
measure the tensor magnetic polarizability of the deuteron in
storage ring experiments. We also confirm an existence of the
effect found by Baryshevsky and Gurinovich, hep-ph/0506135 and
Baryshevsky, hep-ph/0510158; hep-ph/0603191 that the tensor
magnetic polarizability of the deuteron causes the spin rotation
with two frequencies and experiences beating for polarized
deuteron beams in storage rings.
\end{abstract}


\maketitle

\section{Introduction}

Tensor electric and magnetic polarizabilities defined by spin
interactions of nucleons are important parameters of the deuteron
and other nuclei. In particular, measurement of tensor
polarizabilities of the deuteron gives important information about
spin-dependent nuclear forces.

For vector-polarized deuteron beams in storage rings, the main
effects caused by the tensor polarizabilities have been
investigated by Baryshevsky and Gurinovich \cite{Bar3} and
Baryshevsky \cite{Bar4}. The tensor magnetic polarizability,
$\beta_T$, produces the spin rotation with two
frequencies instead of one, 
beating with a frequency proportional to $\beta_T$, and causes
transitions between vector and tensor polarizations
\cite{Bar3,Bar4}. We confirm the existence of these effects and
carry out a detailed calculation of deuteron spin dynamics in
storage rings. We use the matrix Hamiltonian obtained in Ref.
\cite{PRC} and derive general formulae describing the evolution of
the spin. 
We show that an initially tensor-polarized deuteron beam can
acquire a final horizontal vector polarization of the order of
1\%. This effect makes it possible to measure the tensor magnetic
polarizability of the deuteron in storage rings.

The system of units $\hbar=c=1$ is used.

\section{Hamiltonian
approach in the method of spin amplitudes}

The method of spin amplitudes uses quantum mechanics formalism to
more easily describe spin dynamics (see Ref. \cite{CRMZ}). Vector
and tensor polarizations of particles and nuclei with spin $S\ge1$
are specified by the unit polarization vector $\bm P$ and the
polarization tensor $P_{ij}$, which are given by \cite{MShY}
\begin{eqnarray}
P_i \!=\!\frac{<S_i>}{S}, ~~ 
P_{ij}\! =\! \frac{3 <S_iS_j\! +\!
S_jS_i>\!-2S(S\!+\!1)\delta_{ij}}{2S(2S\! -\!
1)}, \label{eq1}\end{eqnarray} where $S_i$ are 
spin matrices and $i,j=x,y,z$. 
Additional tensors composed of products of three or more spin
matrices are needed only for the exhaustive description of
polarization of particles and nuclei with spin $S\ge3/2$.


The nontrivial spin dynamics predicted in Refs. \cite{Bar3,Bar4},
produced by the tensor electric and magnetic polarizabilities of
the deuteron, are a good example of the importance of spin-tensor
interactions in the physics of polarized beams in storage rings.
To describe tensor interactions of the deuteron with the method of
spin amplitudes, one should use three-component spinors and
3$\times$3 matrices. The method of spin amplitudes is
mathematically advantageous because transporting the
three-component spinor is much simpler than transporting the
three-dimensional polarization vector $\bm P$ and five independent
components of the polarization tensor $P_{ij}$ together.

We follow the traditional quantum mechanical approach \cite{RF}
and use the matrix Hamilton equation and the Hamilton operator
${\cal H}$ for determining an evolution of the spin wave function:
\begin{equation}
 \begin{array}{c}
 i \frac{d\Psi}{dt}=H\Psi, ~~~ \Psi=\left(\begin{array}{c}
C_{1}(t)
 \\ C_{0}(t) \\ C_{-1}(t) \end{array}\right),
 ~~~ H_{ij}=<i|{\cal H}|j>,
 \end{array}\label{eq19t}\end{equation}
where $H$ is $3\times3$ matrix, $H_{ij}$ are matrix elements of
${\cal H}$, $\Psi$ is the three-component spin
wave function (spinor), with the wave function $|i>$ defined by $C_i=1$, 
and $i,j=1,0,-1$. In the case considered, the Hamilton operator is
given by \cite{PRC}
\begin{equation} \begin{array}{c}
{\cal H}={\cal H}_0+\bm S\cdot\bm\omega_a-(\alpha_T\beta^2
S^2_\rho+\beta_T S_z^2)B^2_z\gamma,
\end{array} \label{eqapp} \end{equation}
where $\bm\omega_a$ is the angular velocity of spin 
precession relative to the direction of momentum (g$-$2
precession), $\alpha_T$
is the tensor electric polarizability, $\beta=v/c$ is the
normalized velocity, $B_z$ is the vertical magnetic field, and
$\gamma$ is the Lorentz factor.

A determination of spin dynamics can be divided into several
stages, namely,
(i) a solution of Hamilton equation (\ref{eq19t}) and a
determination of eigenvalues and eigenvectors of the matrix
Hamiltonian $H$,
(ii) a derivation of the spin wave function consisting in a
solution of a set of three linear algebraic equations,
and (iii) a calculation of the time evolution of the polarization
vector and the polarization tensor.

\section{Dynamics of deuteron spin in storage rings}

Corrections to the Hamilton operator for the deuteron
polarizabilities contain scalar and tensor parts. The scalar part
is spin-independent and can be disregarded. The general form of
the matrix Hamiltonian $H$ has been found in Ref. \cite{PRC}. When
we consider the deuteron spin dynamics in a uniform magnetic
field, the matrix Hamiltonian is Hermitian and takes the form
\begin{equation}
 \begin{array}{c}
 H=\left(\begin{array}{ccc} E_{0}\!+\!\omega_0\!+\!{\cal A}\!+\!{\cal B} & 0 & {\cal A} \\
0 & E_{0}\!+\!2{\cal A} & 0 \\
{\cal A} & 0 & E_{0}\!-\!\omega_0\!+\!{\cal A}\!+\!{\cal B}
\end{array}\right),
\end{array}\label{eqMH}\end{equation}
where \cite{PRC}
\begin{equation}
\begin{array}{c}
{\cal A}=-\frac{1}{2}\alpha_TB_z^2\gamma\beta^2, ~~~ {\cal
B}=-\beta_TB_z^2\gamma.
\end{array}\label{eqMHt}\end{equation}
$\omega_0=\left(\omega_a\right)_z$ is the angular frequency of
spin rotation (g$-$2 frequency) and $E_0$ is the zero energy
level. The nondiagonal components in Eq. (\ref{eqMH}) are
nonresonant and can be disregarded because their average effect on
the rotating spin is zero. Equation \ref{eqMH}) shows that the
deuteron spin in the horizontal plane is affected not only by the
tensor magnetic polarizability but also by the tensor electric
one.

It can be easily checked that matrix Hamiltonian (\ref{eqMH})
coincides with Hamilton operator (\ref{eqapp}) expressed in matrix
form. The coincidence of Eqs. (\ref{eqapp}) and (\ref{eqMH})
results from the fact that the considered Hamilton operator is
independent of coordinates.

Calculations of spin dynamics defined by Eqs.
(\ref{eqMH}),(\ref{eqMHt}) are much simpler than the calculations
fulfilled in Ref. \cite{PRC} because of the absence of resonance
effects. The connection between spin amplitudes and components of
polarization vector and polarization tensor is given by Eq. (44)
in Ref. \cite{PRC}. If the deuteron beam is vector-polarized and
the direction of its polarization is characterized by the
spherical angles $\theta$ and $\psi$, the general equation
defining the evolution of deuteron polarization has the form
\begin{equation}
\begin{array}{c}
P_\rho(t)=\sin{\theta}\cos{(\omega_0t+\psi)}\cos{(bt)}\nonumber\\
-\sin{\theta}\cos{\theta}\sin{(\omega_0t+\psi)}\sin{(bt)},\nonumber\\
P_\phi(t)=\sin{\theta}\sin{(\omega_0t+\psi)}\cos{(bt)}\nonumber\\+\sin{\theta}\cos{\theta}\cos{(\omega_0t+\psi)}\sin{(bt)},
~~~ P_{z}(t)=P_{z}(0),
\end{array}
\label{prop}
\end{equation}
where
\begin{equation}
b={\cal B}-{\cal
A}=-\left(\beta_T-\frac{1}{2}\alpha_T\beta^2\right)B_z^2\gamma.
\label{eqb}
\end{equation}
The initial vector polarization is given by $$\bm
P(0)=\sin{\theta}\cos{\psi}\bm e_\rho+\sin{\theta}\sin{\psi}\bm
e_\phi+\cos{\theta}\bm e_z.$$

Eqs. (\ref{prop}),(\ref{eqb}) show that the tensor
polarizabilities cause the spin rotation of the vector-polarized
deuteron beam with two frequencies $\omega_0\pm b$ instead of
$\omega_0$ and therefore experience beating with the frequency
$\Delta\omega=-2b$. Thus, these equations confirm the conclusion
given by Baryshevsky and Gurinovich \cite{Bar3} and Baryshevsky
\cite{Bar4}. Eq. (\ref{eqb}) displays that the spin rotation is
also affected by the tensor electric polarizability. Although the
effect predicted in Refs. \cite{Bar3,Bar4} is not negligible, its
direct observation is very difficult. There are three independent
theoretical predictions for the value of the tensor electric
polarizability of the deuteron, namely
$\alpha_T=-6.2\times10^{-41}$ cm$^3$ \cite{CGS},
$-6.8\times10^{-41}$ cm$^3$ \cite{JL}, and $3.2\times10^{-41}$
cm$^3$ \cite{FP}. The first two values are very close to each
other but they do not agree with the last result. The theoretical
estimate for the tensor magnetic polarizability of deuteron is
$\beta_T=1.95\times10^{-40}$ cm$^3$ \cite{CGS,JL}. The duration of
measurement $t$ is restricted by the spin coherence time $\tau$.
If we base our estimate on the values corresponding to the planned
deuteron electric-dipole-moment experiment in storage rings
\cite{OMS} ($\gamma=1.28,~\beta=0.625,~\tau\sim1000$ s, and
$B_z=3$ T), then $b\sim10^{-5}$ s$^{-1}$ and $bt\lesssim10^{-2}$.
In this case, systematical errors caused by betatron oscillations,
field defects, and misalignments of magnets can appreciably exceed
small perturbations of spin rotation produced by the deuteron
tensor polarizabilities.

We propose a significant improvement in the precision of a
possible
experiment. Measurement 
of the effect 
can be strongly simplified with the use of a
\emph{tensor-polarized} deuteron beam. If the initial vector
polarization of such a beam is zero, any interactions of the
magnetic moment of deuteron with external fields cannot lead to
the appearance of vector polarization. Therefore, nonzero vector
polarization of the beam can be produced by nothing but the tensor
interactions. The initial tensor polarization can correspond to a
zero projection of the deuteron spin onto the preferential
direction. When this direction is defined by the spherical angles
$\theta$ and $\psi$, the initial polarization is given by
\begin{equation}\begin{array}{c}
\bm P(0)=0, ~~~
P_{\rho\rho}(0)=1-3\sin^2{\theta}\cos^2{\psi}, \\
P_{\phi\phi}(0)=1-3\sin^2{\theta}\sin^2{\psi}, ~~~
P_{zz}(0)=1-3\cos^2{\theta}, \\ 
P_{\rho\phi}(0)=-\frac32\sin^2{\theta}\sin{(2\psi)}, \\ P_{\rho
z}(0)\!=\!-\frac32\sin{(2\theta)}\cos{\psi}, ~~~ P_{\phi
z}(0)\!=\!-\frac32\sin{(2\theta)}\sin{\psi}.
\end{array}\label{intvi}
\end{equation}
In this case, the time dependence of the polarization vector has
the form
\begin{equation}
 \begin{array}{c}
P_\rho(t)=\sin{(2\theta)}\sin{(\omega_0t+\psi)}\sin{(bt)},
\nonumber\\
P_\phi(t)=-\sin{(2\theta)}\cos{(\omega_0t+\psi)}\sin{(bt)},
~~~ 
P_{z}(t)=0.
\end{array}
\label{etpwi}
\end{equation}
The final vector polarization is horizontal.

Spin dynamics can be easily calculated for any other initial
tensor polarization of the deuteron beam.


Eq. (\ref{etpwi}) shows the possibility of measurement of the
quantity $b$ in storage ring experiments. The final vector
polarization of the beam is of the order of $1\%$.

\section{Discussion and summary}

The analysis presented confirms the results obtained by
Baryshevsky and Gurinovich \cite{Bar3} and Baryshevsky
\cite{Bar4}. It is shown that the predicted rotation of the
deuteron spin with two frequencies, beating, and transitions
between vector and tensor polarizations in a uniform magnetic
field are produced not only by the tensor magnetic polarizability
but also by the tensor electric polarizability. Nevertheless, the
latter quantity gives a minor contribution to the effect. If
experimental conditions correspond to the planned deuteron
electric-dipole-moment experiment in storage rings \cite{OMS}
($\beta^2=0.4$) and theoretical estimates for the tensor
polarizabilities of the deuteron given in Refs. \cite{CGS,JL,FP}
are used, the expected relative importance of $\beta_T$ is one
order of magnitude greater.

Unfortunately, the expected spin coherence time (about 1000 s
\cite{OMS}) is too short to register beating. In this case,
systematical errors can prevent the observation of small
perturbations of spin rotation conditioned by the tensor
polarizabilities of the deuteron. Possibly, the effect predicted
in Refs. \cite{Bar3,Bar4} may be discovered with the use of a
Penning trap used as a minicyclotron, since the Penning trap
provides a much longer duration of measurement.

The precision of a possible storage ring experiment can be
significantly improved, if the deuteron beam is 
tensor-polarized. If the initial vector polarization of such a
beam is zero, any interactions 
linear in the spin cannot lead to the appearance of vector
polarization. The final vector polarization also cannot result
from the betatron oscillations, field defects, misalignments of
magnets and other potential sources of systematical errors.
Therefore, the tensor interaction of spin with the magnetic field
defined by Eqs. (\ref{eqMH}),(\ref{eqMHt}), and
(\ref{eqb})--(\ref{etpwi}) is the only reason for nonzero vector
polarization of the beam.
Eq. (\ref{etpwi}) shows that the beam can acquire the final
horizontal vector polarization of the order of 1\%. Known
experimental methods \cite{measure} permit safe measurement of
such a polarization and therefore a determination of the tensor
magnetic polarizability of the deuteron. In addition, the
observation of the predicted effect would prove the importance of
taking into account spin-tensor interactions in storage ring
physics.

The rotation of the polarization vector in the horizontal plane
complicates a measurement of beam polarization. Two possible
solutions of this problem are to eliminate the g$-$2 spin rotation
with an additional radial electric field (see Ref. \cite{EDM}) or
to use Siberian snakes and a stimulation of spin resonance
\cite{Bar3,Bar4}. Both of these methods need additional devices
(electric field plates or resonators) and result in a significant
complication of an experimental setup. We propose direct
measurement of the beam polarization in a fixed direction. The
availability of the proposed method is based on the attainability
of the spin coherence time about 1000 s. In this case, one holds
the predicted polarization of the beam until the appearance of a
measurable vector polarization. Spin coherence can be maintained
because of the independence of the g$-$2 frequency on the deuteron
momentum for an appropriate length of straight ring sections
\cite{OMS}. The spin coherence may also be supported by a
small 
deuteron momentum spread $\Delta p/p$. It is 
important that the ratio of the g$-$2 frequency to the cyclotron
one is small enough
[$\omega_{0}/\omega_c=\gamma(g-2)/2=-0.143\gamma$ in semicircular
sections]. Under these conditions, the spin evolution is
clearly defined 
and one can choose an appropriate measurement time interval. The
experiment can be implemented in one of the existing storage
rings.

Since Eq. (\ref{eqb}) contains both tensor polarizabilities, the
tensor electric polarizability of the deuteron can also be
measured in the proposed experiment and the obtained precision of
polarization measurements \cite{measure} should be significantly
improved. The experiment should be
performed with 
different values of the beam momentum (and the vertical magnetic
field) on the same ring. In this case, both tensor
polarizabilities of the deuteron can be determined. Another method
of determination of the deuteron's tensor electric polarizability
has been proposed in Ref. \cite{PRC}.
The use of the Penning trap in experiments with tensor-polarized
deuterons can also be helpful for measuring the tensor
polarizabilities.

\section*{Acknowledgements}

The author is grateful to V.G. Baryshevsky for bringing the
considered problem to his attention and for helpful discussions
and comments. This work was supported by the Belarusian Republican
Foundation for Fundamental Research.

\label{last}
\end{document}